\documentclass[useAMS,usenatbib]{mn2e}

\usepackage{graphicx}
\usepackage{mathpazo}
\usepackage{amsmath}	
\usepackage{amssymb}

\usepackage{tabulary}

\newcommand{\ignore}[1]{}
\newcommand{\be}{\begin{equation}} \newcommand{\ee}{\end{equation}}

\def\ba#1\ea{\begin{align}#1\end{align}}
\newcommand{\bit}{\begin{itemize}}
\newcommand{\eit}{\end{itemize}}

\newcommand{\nn}{\nonumber}  \newcommand{\ra}{\rightarrow}

\renewcommand{\a}{\alpha} \renewcommand{\b}{\beta}

\def\slashb#1{\setbox0=\hbox{$#1$}#1\hskip-\wd0\dimen0=5pt\advance
        \dimen0 by-\ht0\advance\dimen0 by\dp0\lower0.5\dimen0\hbox
          to\wd0{\hss\sl/\/\hss}}

\title[A Redshift-Dependent Color-Luminosity Relation in Type 1a Supernovae]{A Redshift-Dependent Color-Luminosity Relation in Type 1a Supernovae}
\author[Gopolang M. Mohlabeng and John P. Ralston]{Gopolang M. Mohlabeng\thanks{E-mail:
mohlabeng319@yahoo.com; ralston@ku.edu} and John P. Ralston\\
Department of Physics and Astronomy, University of Kansas, Lawrence, KS, 66045}

\begin{document}

\date{}

%\pagerange{\pageref{firstpage}--\pageref{lastpage}} \pubyear{2002}

\maketitle

\label{firstpage}

\begin{abstract}

Type 1a supernova magnitudes are used to fit cosmological parameters under the assumption the model will fit the observed redshift dependence. We test this assumption with the Union 
2.1 compilation of 580 sources. Several independent tests find the existing model fails to account for a significant correlation of supernova color and redshift.  The correlation of 
magnitude residuals relative to the $\Lambda CDM$ model and $color \times redshift$ has a significance equivalent to 13 standard deviations, as evaluated by randomly shuffling the 
data. Extending the existing $B-V$ color correction to a relation linear in redshift improves the goodness of fit $\chi^{2}$ by more than 50 units, an equivalent 7-$\sigma$ 
significance, while adding only one parameter. The $color-redshift$ correlation is quite robust, cannot be attributed to outliers, and passes several tests of consistency. We review 
previous hints of redshift dependence in color parameters found in bin-by-bin fits interpreted as parameter bias. We show that neither the bias nor the change $\Delta \chi^{2}$ of our 
study can be explained by those effects. The previously known relation that bluer supernovae have larger absolute luminosity tends to empirically flatten out with increasing redshift. 
The best-fit cosmological dark energy density parameter is revised from $ \Omega_{\Lambda} =0.71 \pm 0.02$ to $ \Omega_{\Lambda} = 0.74 \pm 0.02$ assuming a flat universe. One possible 
physical interpretation is that supernovae or their environments evolve significantly with increasing redshift.

\end{abstract}


\section{Introduction}
Observations of Type 1a supernovae ($Sn1a$) provide evidence for an accelerating expansion of the universe and dark energy. Recent surveys have accumulated sufficient data 
that statistical uncertainties may be smaller than systematic ones. Supernovae are imperfect standard candles, and systematic corrections for their absolute luminosity have evolved 
over time. Besides a correction for time-stretch \Citep{Phillips1993} there is now a color ($c$) correction parameter \Citep{VanDenBurg1995,Tripp1998} called $\b$. Fits to 
the cosmological model replace the observed magnitude $m_{B}$ by $m_{B} -\b c$, representing the empirical fact that $Sn1a$ with bluer colors tend to be intrinsically brighter. 

After such redshift-independent corrections, cosmological parameters are fit using Type 1a supernova data under a hypothesis that the cosmological model describes the data. Despite 
the existence of publicly-accessible $Sn1a$ data, we find few independent tests of the assumption. As noted by Vishwakarma (2011), for the past few years studies with 
high statistics have focused on parameter estimation using methods that are not designed to test the model.

Here we pose simple and direct hypotheses tests comparing different models of $Sn1a$ magnitude corrections. In the existing null model the correction parameters 
$\theta_{i} =(\a, \, \beta, \delta)$ for stretch, color and galaxy type have no redshift dependence. We compare that to a model using a two term Taylor expansion 
$\theta_{i}  \ra \theta_{i}(z)$. For the color parameter, the expansion is 
\ba 
\b c \ra \b(z) c=\b_{0}c+\b_{1} (c z-\overline{c z}).  \label{colorz} 
\ea 

We find the null model $\b_{1} \equiv 0$ fails with high statistical significance. We call the empirical correlation the ``$color-redshift$ effect.'' It is related to previous hints 
of $SN1a$ evolution with redshift discussed below.

\subsection{Assumptions and Definitions}

Before proceeding we consider Malmquist (magnitude) selection bias and ``other'' selection bias. The bias that observing higher redshift $Sn1a$ tends to select a brighter 
population is well known. A proper test is not concerned with the absolute magnitudes directly, but instead with the {\it differences} between the systematically corrected data and 
the cosmological model. When we extend parameters $\theta_{i} \ra \theta_{i}(z)$ it stands on the same testable footing as testing redshift dependence of the dark matter and dark 
energy density parameters. Moreover, while a population bias may still contribute to the ``lever arm'' of fitting parameters, it should not make a true model fitting the data appear 
to be false. Regarding other, unspecified ``bias'' as explanation for the effects we find: We regard that as fair game. When bias exists it needs to be explored, while the business of 
testing hypotheses is itself neutral on the interpretation.

We use the 580 points of the Union 2.1 compilation (Suzuki et al. 2012, henceforth S12) to test the current model. The data's distance moduli $\mu_{B}$ corrected for color ($c$), stretch  ($x_1$) and a 
certain probability $P_{m}= P(m_{*}^{host} \,< m_{*}^{Theshold})$ of the host galaxy are
\ba 
\mu_{B}(\alpha, \, \beta,\, \delta, \, M_{B}) = m_B + \alpha\, x_{1} - \beta \, c + \delta\, P_{m}  - M_{B}. \label{mbdef}
\ea 
Here $x_{1} = s - \bar s$, where $s$ is the time stretch factor with redshift ($z$) corrections applied \Citep{Goldhaber, Guy2007}. Symbol $c = color - \overline{color}$, where 
$color = (B - V)_{max} + 0.057$ and the Johnson-Cousins $B$ stands for blue and $V$ the visual magnitude. Overbars denote mean values, which have also been subtracted from $P_{m}$ to remove a degeneracy with the value of $M_{B}$.

As mentioned earlier, we fit $\a(z)$, $\b(z)$ and $\delta(z)$ to two-term polynomials. We believe that transparent and reproducible data analysis is of intrinsic value, 
and deliberately avoid statistically elaborate methods. Nevertheless we have reproduced the analysis elements of our main references which are reproducible with the data that has been 
published. The comparison of our independent study with those produced with data compendia focused on cosmological parameter estimation should be a reason to extend those analyses to 
include redshift dependent parameters.

Models assume a $\Lambda$CDM cosmology with zero radiation density, dark energy density $\Omega_{\Lambda}$ and dark matter density $\Omega_{m}$. 
The predicted luminosity distance is \Citep{weinberg} :  
\ba
 d_{L}(z) &= \frac{c(1+z)}{H_{0} \sqrt{\Omega_{k}} }sinh \left( \sqrt{\Omega_{k}} \int_{1/(1+z)}^{1} \frac{dx}{x^{2}H(x) } \right) ; \nonumber \\ &  \:\:\:\:\: where \:\:\:\:\: 
H(x)=\sqrt{\Omega_{\Lambda}x^{-3(1+w)} + \Omega_{m}x^{-3} + \Omega_{k}x^{-2}}.
\ea
Here $H_0=70 km s^{-1} Mpc^{-1}$ is the Hubble constant, which is degenerate in fits with $M_{B}$. The flat cosmology has $\Omega_{k}=1-\Omega_{m}-\Omega_{\Lambda}=0$. We use the 
standard value $w=-1$; little sensitivity of $w$ to the new correlation appeared in preliminary work. The distance modulus $ \mu_{model}$ is defined by:
\ba
 \mu_{model}(z,\Omega_{\Lambda},\Omega_{m}) = 5\log_{10}( {d_{L}(z) \over Mpc} ) + 25.
\ea 

\section{ Data and Analysis} 
\label{sec:correl}

We began with the 2008 Union 1 compilation (Kowalski et al. 2008, henceforth K08) before our main data set had been released. The data tables reported only magnitude uncertainties, 
omitting the systematic errors, which are much larger. We found a large correlation of $color \times z$ and the residuals of the 307 point `3$\sigma$ set'' defined in the paper. One 
$color \times redshift$ parameter $\beta_{1}$ also yielded a difference $\Delta_{mag}\chi^2 >30$, using the raw uncertainties. Next we studied the Union 2.0 compilation of 557 
$Sn1a$ (Amanullah et al. 2010, henceforth A10), which gave both raw and total uncertainties. The relative improvement of $\chi^{2}/df$ using $\b_{1}$ was found to be quite comparable. 
Finally the Union 2.1 compilation of S12 gave enough detail to reproduce its $\chi^{2}/df$ using its best-fit parameters 
$\a, \, \b, \delta, \Omega_{m}, \,M_{B}$ and $w=-1$. Our results here are confined to the ``3$\sigma$ set'' of Union 2.1 data defined by S12.  

In all work, data reversion and analysis were done twice, by independently written codes comparing outputs while sharing no common elements. 

\begin{figure}
\centering
\includegraphics[width=3.5in,height=2.5in]{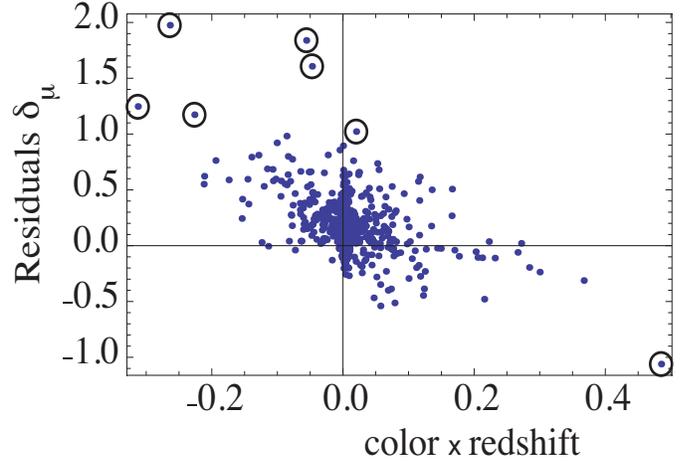}
\caption{ \small Distance modulus residuals $\delta_{\mu} = \mu_{B} -\mu_{model}$ relative to the best-fit $\Lambda$CDM model versus $color \times redshift$. Calculation assumes a 
flat universe and $w=-1$. The Pearson correlation coefficient of $r= -0.52$ has the significance of a $13 \sigma$ effect, as evaluated by a simulation shuffling the data randomly. Seven 
points discussed in the text are indicated by circles.  Data is the $3\sigma$ set of the Union 2.1 compilation.}
\label{fig:suzukires}
\end{figure}

\subsection{ Correlation Results} 

Residuals $\delta_{\mu}$ are the differences between the distance moduli and the model: 
\ba 
\delta_{\mu} = \mu_{B} -\mu_{model}. 
\ea  
Residuals are computed using the top line of Table 6 in S12, except we impose $w=-1$. The correlation of residuals with $color \times redshift$ is readily seen by eye in Figure 
\ref{fig:suzukires}. Notice it extends over both positive (more red) and negative (more blue) color. 

The Pearson correlation coefficient $r$ quantifies the correlation:
\ba
r(x,y) = \frac{\sum_{i}(x_{i} - \bar x)(y_{i} - \bar y)}{\sqrt{\sum_{i}(x_{i} - \bar x)^{2}}\sqrt{\sum_{i}(y_{i} - \bar y)^{2}}} 
\ea
We find $r_{SN}=r(\delta_{\mu}, \, c\ z) = -0.52$. We estimated the significance of $r_{SN} $ with two Monte-Carlo simulations. We randomly shuffled the $(z_{i}, c_{i})$ data elements 
and recalculated 10,000 random correlations $r_{random}$. The distribution of $r_{random}$ was consistent with a Gaussian with standard deviation $\sigma_{r-random} = 0.042$ and mean 
$\sim \pm 0.001$.  Numerically indistinguishable results were found when the data elements were combined into linear combinations $(z'_{i}, c'_{i})$ preserving the data's covariance 
matrix between redshift and color. This was done to verify that the substantial correlation (equalling -0.22) between redshift and color is not an issue with the correlation of 
residuals we observe. In both cases the data's correlation of $r_{SN}$ is about $13\sigma_{r-random}$ from the mean. The estimated $P$-value (of order $10^{-39}$) is too small to 
simulate or interpret as a fluctuation. This correlation shows that the data has a redshift of dependence that is not predicted by the model producing the residuals.

One might ask whether outliers produce the correlation. Actually outliers were already removed in the Union 2.1 data. Starting with 753 $Sn1a$, data differing more than $3 \sigma$ 
from the model were discarded. The detailed cuts listed in S12 are (1) requiring a $CMB$-centric redshift greater than 0.015; (2) requiring at least one point between -15 and 6 
rest-frame days from B-band maximum light; (3) requiring at least five valid data points; (4) requiring the entire 68\% confidence interval for $x_{1}$ between -5 and +5;
(5) requiring at least two bands with rest-frame central wavelength coverage between 2900 $A^{o}$ and 7000 $A^{o}$; (6) requiring at least one band be redder than rest-frame U-band 
(4000 $A^{o}$). We note the extra attention to color is more demanding than many previous studies. Using $w=-1$ there are seven points indicated with circles in Figure 
\ref{fig:suzukires} with $|\delta_{\mu}|> 3\sigma$ for the residuals shown. Removing those points reduced the correlation to $r_{SN, \, cut}= -0.45$, which equals $11\sigma$.

\subsection{A Model with One New Parameter:} We define our first fit statistic $\chi^{2}$ as
\ba
 \chi^{2} = \sum_{SNe} \, \frac{(\mu_{B}(\a ,\b, \delta, M_{B}) - \mu_{model}(z, \Omega_{M}, \Omega_{\Lambda}, w))^{2}}{\sigma_{\mu}^{2}}. \label{chi} 
\ea
Here $\sigma_{\mu}^{2}$ are the complete (statistical and systematic) errors from the $SCP$ website \footnote{See $http://supernova.lbl.gov$}. 
The null model of a redshift-independent color correction gives the best-fit $\chi^2 = 550$ for $580-5$ degrees of freedom ($df$) (Table 1). 

We computed the change $\Delta \chi^2$ after extending $\a$ and $\delta$ to a Taylor series linear in $z$:   
 \ba
(fixing\:\beta_{1}=\delta_{1}=0):  \:\:\: &\alpha  \ra \alpha_{0}+\alpha_{1} z;  \ \:\:\: \a_{1}=-0.036 ; \:\Delta \chi^{2} = 3.8;\nn  \\
 (fixing\: \alpha_{1}=\beta_{1}=0):  \:\:\: & \delta \ra \delta_{0}+\delta_{1} z;  \ \:\:\: \delta_{1}= -0.093; \: \Delta \chi^{2} = 2. \nn
\ea 
 
Our reason for varying parameters one at a time was computational simplicity. These values do not indicate a highly significant redshift dependence of $\a$ or $\delta$.
However $\Delta \chi^{2} = 3.8$ slightly exceeds an equivalent $2 \sigma$ confidence level disfavoring a constant stretch parameter.

On the other hand varying $\beta_{1}$ (with $\a_{1}=\delta_{1}=0$) produced a best fit with $\chi^2 =500$ for $(580-6)df$. That is a 50 unit decrease in $\chi^2$ from the addition of 
one parameter: see Table 1. When a null hypothesis is extended by one parameter, Wilk's theorem predicts that $\Delta \chi^{2}$ will be distributed by the $\chi_{1}^{2}$ distribution, 
and be typically of order one. On that basis 50 units of $\chi^{2}$ is equivalent to a Gaussian fluctuation of $7.1 \sigma$, which has a chance probability of order $10^{-12}$. This 
rules out the null model on an independent basis.

The low value of $\chi^2/df \lesssim 1$ is unlikely on a statistical basis. A simple explanation exists. The data compilations assign systematic errors with a step 
$\sigma_{\mu}^{2} \ra \sigma_{\mu}^{2}+\sigma_{int}^{2}$. Symbol $\sigma_{int}^{2}$ is an intrinsic scatter parameter adjusted to bring $\chi^2/df$ of each sample to unity. 
(See K08, A10, and especially S12 following Eq. \ref{chi}). After $\b_{1}$ has been fit the same procedure could re-adjust $\sigma_{int}^{2}$ to give $\chi^2/df \ra 1$. Except for one 
test below, we make no adjustment or variation of errors, accepting the reported values. We also note the absolute magnitude $M_{B}$, as well as $\a$, $\b$, etc. have been called 
``nuisance parameters'' that are not reported in many papers. We believe fitted parameters have physical importance, and Table 1 reports everything needed to reproduce our results.  

\begin{table*}
 \centering
 \begin{minipage}{160mm}
\caption{\small  Comparison of fits without accounting for the $color-redshift$ effect ($\beta_{1}=0$) and including it ($\beta_{1}\neq 0$). Parameters held fixed are indicated by an 
asterisk. The $color-redshift$ effect produces a highly significant improvement of the fit even with unphysical constraints such as $\Omega_{m}=\Omega_{\Lambda}=0$. $\chi^2$ is defined
by Eqs. \ref{chi} and \ref{mbdef}, with $\sigma_{\mu}^{2}$ the distance modulus uncertainties from the data tables, except for the last line (indicated by $\dagger$) using 
parametrically-fit errors of the type found in current literature. $\chi_{min}^{2}$ and $\Delta \chi^{2}$ have been rounded to the nearest whole number.}
  \begin{tabular}{cccccccccc}
  \hline
$\Omega_{m}$ &  $\Omega_{\Lambda}$ & $\alpha_{0}$ & $\beta_{0}$ & $\delta_{0}$ & $M_{B}$ &  $\beta_{1}$ & $\chi_{min}^{2}$ & $\Delta \chi^{2}$ \\  
\hline  
0.291 $\pm$ 0.022 &1-$\Omega_{m}$ & 0.105 $\pm$ 0.007 & 2.31 $\pm$ 0.05 & -0.022 $\pm$ 0.03 & -19.133 $\pm$ 0.013 & 0$^{*}$ & 550 & 0  \\ 
0$^{*}$ &  0$^{*}$  & 0.10 $\pm$ 0.006& 2.51 $\pm$ 0.07 & - 0.11 $\pm$ 0.024 & -19.1 $\pm$ 0.012 & -1.35 $\pm$ 0.22& 534 & 16   \\
0$^{*}$ & 0.31 $\pm$ 0.003 & 0.105 $\pm$ 0.00004& 2.61 $\pm$ 0.005& -0.054 $\pm$ 0.0007&-19.05 $\pm$ 0.0002& -1.59 $\pm$ 0.051& 511 & 39 \\
0.260 $\pm$ 0.021 &1-$\Omega_{m}$ & 0.102 $\pm$ 0.007 & 2.62 $\pm$ 0.07 & -0.038 $\pm$ 0.03 &-19.14 $\pm$ 0.013 & -1.61 $\pm$ 0.23 & 500 & 50\\ 
0.259 $\pm$ 0.07  & 0.737 $\pm$ 0.13 & 0.102 $\pm$ 0.0065 & 2.62 $\pm$ 0.07 & -0.038 $\pm$ 0.027 &-19.16 $\pm$ 0.073 & -1.61 $\pm$ 0.227 & 500 & 50   \\
$0.236 \pm 0.021$ & 1-$\Omega_{m} $ & 0.11 $\pm$ 0.0085   & 2.91 $\pm$ 0.074 & -0.05 $\pm$ 0.029  &   -19.15 $\pm$ 0.014 & -2.40 $\pm$ 0.21  &  464 &  $109^{\dagger} $  \\

 \hline
\end{tabular}
\end{minipage}
\label{tab:firstable}
\end{table*}

\begin{figure}
\centering
\includegraphics[width=3in,height=2in]{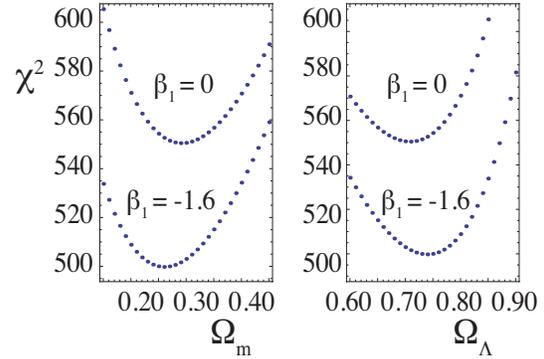}
\caption{ \small  Left panel: $\chi^{2}$ versus $\Omega_{m}$ without $color \times redshift$ parameter ($\beta_{1} =0$, upper points) and including it ($\beta_{1}=-1.6$, lower points). 
Right panel: Same as left with $\Omega_{\Lambda}$ on the $x$-axis. Fits use $\Omega_{m}+\Omega_{\Lambda}=1$ and best-fit values of the remaining parameters point-by-point. }
\label{fig:chiplot.eps}
\end{figure}

\begin{figure}
\centering
\includegraphics[width=3in]{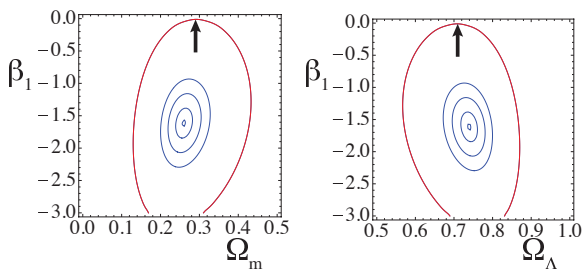}
\caption{ \small  {\it Left Panel:} Contours of constant $\chi^{2}$ in the $(\Omega_{m}, \, \beta_{1})$ plane. Inner contours are $\chi_{min}^{2} =500+j^{2}$ for $j=$0, 1, 2, 3, corresponding to 1, 2, 3 units of Gaussian confidence levels $\sigma$. Outer contour (red online) is $\chi^{2}=550$, (7.1 $\sigma$ equivalent), which intersects the null model $(\beta_{1}=0, \Omega_{m}=0.29)$, as indicated by the small arrow. {\it Right panel:} Same as left 
using the $(\Omega_{\Lambda}, \, \beta_{1})$ plane with the outermost contour intersecting at $(\beta_{1}=0, \,\Omega_{\Lambda}=0.71)$. Fits use $\Omega_{m}+\Omega_{\Lambda}=1$ and best-fit 
values of the remaining parameters point-by-point.}
\label{fig:Beta1OmAndOmlaContours.eps}
\end{figure}

Figures \ref{fig:chiplot.eps} and \ref{fig:Beta1OmAndOmlaContours.eps} show $\chi^{2}$ versus $\Omega_{m}$ and $\Omega_{\Lambda}$ with and without accounting for the $color-redshift$ 
effect. The plots use $\a_{0}\, \beta_{0}, \, \delta_{0}, \, M_{B}$ evaluated at their best-fit values point by point and $\Omega_{m}+\Omega_{\Lambda}=1$. As Table 1 shows, the 
important cosmological parameters $\Omega_{m}$ and $\Omega_{\Lambda}$ are sensitive to the value of $\beta_{1}$. The significance of the $color-redshift$ effect for $\Omega_{m}$ and 
$\Omega_{\Lambda}$ depends on how it is assessed. For example, fixing other parameters to the global best fit value \Citep{proceedings} finds that $\Omega_{m}$ and $\Omega_{\Lambda}$ 
shift by more than their 99.95\% (3 $\sigma$) confidence level uncertainties. That is appropriate when other parameters are known. The plots here showing $\chi^{2}$ with other 
parameters floating to their best-fit values are more conservative, and based on the supernova data alone. (They should not be compared with joint confidence intervals from $CMB$ or 
galaxy distribution data.) Following the convention of S12 (and its Table 7) the errors in the Table correspond to $\Delta \chi^{2}=1$. 

\section{Discussion} 

\label{sec:discussion}

The statistical significance of the $color \times redshift$ correlation is sufficiently high that ordinary confidence levels fail to express it. The correlation is very robust and too 
large to be attributed to outliers.
  
Besides the test in Section \ref{sec:correl}, we computed a new difference $\Delta \chi^{2}(N) = \chi^{2}(\beta_{1}=0, \,N)-\chi^{2}(\beta_{1}, \,N)$, 
where $N$ points were selected on the basis of their uncertainty-weighted residuals relative to the null-model fit. For this the data was sorted in order of decreasing 
$\delta_{\mu \, 1}^{2} /\sigma_{1}^{2}> \delta_{\mu, \, 2}^{2}/\sigma_{2}^{2}>...\delta^{2}_{\mu, \, N}/\sigma_{N}^{2}$ as computed from the null-model fit. Discarding the first $N$ 
points and comparing $\beta_{1}=0$ with $\beta_{1}\neq 0$ produces $\Delta \chi^{2}(N)$. This procedure is statistically unfair. Step by step it maximally selects data to confirm the 
null. Its purpose is to explore whether $\Delta \chi^{2}(N)$ might suddenly decrease for some value $N$, signaling a subset with anomalous statistical weight. Instead 
$\Delta \chi^{2}(N)$ decreased smoothly with $N$ from $\Delta\chi^{2}(N=0)=51$ to $\Delta \chi^{2}(N=50)=20.$. Upon reaching $N=50$ the value of $\chi^{2}(\beta_{1}=0, \, N=50)/df$ had 
been artificially decreased from about 550/(550-6) to about 300/(500-6), a relatively large amount.

Two classes of questions naturally arise: \\
{\bf Data Selection, Parameter Estimation Bias, Data Processing:} Selection bias can strongly affect parameter estimation and has received much discussion. Many papers including K08, A10, and S12 extensively discuss the selection of intrinsically brighter supernovae at larger $z$. These papers dismiss (or limit) population bias effects in the statistics of likelihood-ratio tests (minimizing $\chi^{2}$), supposing the model fits the data.

The numerical value of $\b$ has long been a topic of concern. van den Bergh (1995) observed that model calculations showed a bluer-brighter correlation, and suggested the color 
correction of an effective magnitude parameter. Previously Branch and Tamman (1992) noticed that $R_{B}$ found in data fits was smaller than expected from models concerned with dust. 
Yet it is common for best fit-parameters to disagree with ``true'' parameters. Due to that, hints of $SN1a$ evolution previously found were tempered by concerns over parameter bias. 
Astier et al (2006) fit separate $\b$ parameters to selected redshift bins and observed an apparent decrease with $z$. This paper (and all we discuss here) used parametrically-fit 
errors $\sigma_{\mu}^{2}=\sigma_{\mu}^{2}(\a, \, \b)$ as discussed below. Within such a procedure K08 noted a bias in fitting $\b$, as suggested earlier by Wang et al (2006). 
(Kessler et al. 2009, henceforth K09) and Lago et al (2012) also found $\b$ decreasing with $z$ in bin-by-bin fits, and mentioned a possible signal of supernova evolution. After 
finding $\b$ again decreasing bin-by-bin with larger $z$, (Guy et al. 2010, henceforth G10) explored uncertainties in $color$. The result is sometimes described as explaining the 
findings of K09, as in e.g. \citep{Mariner2011, Sullivan2011, Conley2011}. Actually the issues were not cleanly resolved with G10 stating that 
``we are not able to conclude on an evolution of $\beta$ with redshift''. 

Fits to binned data can be suggestive, but they dilute the statistical weight and should include statistical penalties depending on the number of parameters. Our one-parameter approach 
uses no binning, and is able to discriminate between parameter bias and correlation. To illustrate this, we repeated simulations of our data generated from a fixed cosmology with no 
corrections other than $\b c$, plus random noise. Color was generated with random Gaussian color errors with adjustable variance $\sigma_{c}^{2}$. For $z>0.7$ we found 
$\beta_{fit}/\beta_{true} =1/(1+ \sigma_{c}^{2}/0.11^{2})$ closely matched the mean simulation. The difference using the mean $\sigma_{c} =0.091$ of our data subset is $\beta_{true}-\beta_{fit} \sim 1.2$, much like the previous study done by G10. Our entire data set of 580 points gave 
$\beta_{fit}/\beta_{true} =1/(1+ \sigma_{c}^{2}/0.14^{2})$. These results do not depend strongly on the variance of the noise added to $m_{B}$. With the smaller mean $\sigma_{c} =0.052$ of this set we found $\beta_{fit} -\beta_{true}  = 0.39$. That value is comparable to the $\beta_{fit} -\beta_{true} \sim 0.5$ shown for 307 Union points in Fig.5 of K08.   

The actual value of $\b$ is a topic of great interest. Very recently Scolnic et al (2013) directly confront the question of $\beta$-bias, and find that the combination of residual 
scatter due to color and a realistic color distribution will bias $\beta$ by roughly one unit lower than its true value. That work builds on Chotard et al (2011) who suggest that the 
controversy over Type 1a colors can be explained by the dispersion in colors, and by variable features observable in the spectra.

However neither the bias nor the value of $\b$ is relevant to our particular task. The distribution of $\Delta \chi^{2}(\b_{1})$ from the simulations directly tests our methods 
independent of the bias. Our 1000-fit simulations adding random color noise found $\Delta \chi^{2}(\beta_1)$ distributed by $\chi_{1}^{2}$ just as statistical theory predicts. 
(Fitting $\chi_\nu^2$ with $\nu$ a free parameter found $\nu =1\pm O(10^{-2})$). This was found both for the 90 points of $z>0.7$ and the full 580-point data set. The same was found 
when we made color noise increase linearly with $z$. We are not aware of any previous studies that compared hypotheses using $\b \ra \b_{0}+ \b_{1}z$ and global $\Delta \chi^{2}$ fits. 

There is a question whether systematic error assignments might cause a false correlation. Our results for unweighted residuals $\delta_{\mu}$ and $\Delta \chi^{2}/df$ using raw 
magnitude tends to contradict that possibility. We considered adjusting systematic errors to banish the $color-redshift$ effect, but soon realized it would be irresponsible. To explore 
systematic errors in an unbiased way we padded magnitude errors $\sigma_{m_{B}}$ by the rule $(\sigma_{m_{B}})^{2} \ra (\sigma_{\xi}m_{B })^{2}=(\sigma_{m_{B}})^{2} +\xi^{2}$, and 
adjusted parameter $\xi$. The range of $\xi$ spanned the differences of $\overline{\sigma_{ m_{B }}} \sim 0.08$ to $\overline {\sigma_{ \mu_{B}}} \sim 0.22$.
The procedure tests whether points of small error might skew results without unfairly adjusting errors to fit the $FLRW$ model. The recomputed best-fits comparing $\beta_{1} = 0$ and 
$\beta_{1} \neq 0$ yielded a smooth and nearly monotonic variation of $\Delta \chi^{2}(\beta_{1}) =51$ ($\xi=0.001$) to $\Delta \chi^{2} =35.5$ ($\xi=0.2$). The result disfavors 
systematic error assignments causing the correlation. 

We have focused on analysis based on errors as they are published, both for simplicity and so that our results can be reproduced. The recent compendia 
(K08, A10, S12, etc.) compute the errors by parametrically fitting error functions $\sigma_{\mu}^{2}(\a, \, \b, \delta)$ to the cosmological model. The definition for each supernova 
is $\sigma_{\mu}^{2}(\alpha,\beta) = \theta_{i} C_{ij} \theta_{j} + \sigma_{int}^{2}$ where $\theta_{i} = (1,\,\alpha, \, -\beta)$. Here $C_{ij}$ are 580 $3 \times 3$ covariance matrices 
from light curve fitting. The diagonals $C_{ii} = \sigma_{ii}^{2}$ are reported as $(\sigma_{m_{b}}^{2}, \sigma_{x_{1}}^{2}, \sigma_{c}^{2})$. We then followed K09 and Lago et al. (2012) in 
repeating the parametrically fit error calculation using the diagonal covariance matrix elements. Fixing $\b_{1}=0$ we found $\sigma_{int}^{2}=0.016$, as consistent with our references. 
We then compared the best fit with $\b_{1}=0$ with the best-fit $\b_{1}$ to find $\Delta \chi^{2}(\beta_{1})=109$ (Table 1). The value very strongly disfavors the null model. We doubt 
such a large change in $\chi^{2}$ would be due to the (assumed small) unpublished off-diagonal elements of color fitting, but if so it would cast a new light on the entire procedure 
of parametrically-fit errors. The observation of \citep{Lago} that the statistic does not represent a likelihood may be relevant. For completeness, this fit gave 
$\Omega_{m}=0.23 \pm 0.02$. 

{\bf{Physical Interpretation:}} The possibility of $SN1a$ evolving with redshift is well-known. \citep{Tripp1998} wrote that, `` by applying the same type of color correction to 
cosmological supernovae {\it even without knowing whether reddening is intrinsic or due to dust}, one will be able to completely standardize the light output of each explosion...''. 
(Italics are ours.) The current practice of using one ``unique coefficient $\b$...for both dust extinction and any intrinsic colour-magnitude relation'' \citep{Menard2010} allows for 
no evolution. The extraction and interpretation of the color parameter is a very active topic. Besides the work already cited, a significant 
correlation between host galaxy extinction coefficients $A_{V}$ and $\Lambda$CDM residuals suggested evolution or bias, quantified by finding the $\chi^{2}$ value of the ``gold and 
silver'' \citep{Riess2004} data decreased by 23 units of $\chi^{2}$ after varying one new parameter \citep{J&R}. 

A positive value of $\b_{1}$ would increase the redder-is-dimmer correction with increasing $z$, as expected from increasing dust. Instead we find $\b_{1}<0$, a 
decreasing color correction and the opposite effect, which is unexpected and not attributable to conventional dust. The $color-redshift$ effect is, however, consistent with evolution 
of intrinsic luminosity of sources. Extrapolating $\beta(z)=2.62-1.61 z$ naively, the bluer-brighter, redder-dimmer relation would actually reverse for $z \gtrsim 1.6$. An unidentified bias in the observations or data reduction also cannot 
be ruled out. If a bias exists, it is an important issue and hardly a flaw of what we report, which can only be based on the data published. 

In summary, straightforward tests using the Union 2.1 data finds that the current model using constant $\beta$ parameter is ruled out compared to the model 
$\beta(z) =\beta_{0}+\beta_{1}z$. Inasmuch as the fitting of cosmological parameters to Type 1a supernova data hinges on a model called into question, the values and errors of 
those parameters may be questioned. It seems premature to attempt the last word on the highly significant trend we have found.
 
{\textbf{Acknowledgements:}} First results of this study were reported in the {\it South African Institute of Physics (SAIP) Conference, 2012}. We thank
 Eddie Baron, David Branch, Hume Feldman, Greg Rudnick, Bruce Bassett, John Marriner, Graham Wilson and Ned Wright for helpful 
comments. Research supported under DOE Grant Number DE-FG02-04ER14308 and the Fulbright Program.

\end{document}